\documentclass{appolb}
\usepackage{graphicx}
\def\sumint{\int \! \!\ \! \! \! \! \!\ \! \! \!\! \!\sum}
\newcommand{\blf}[1]{\bf  {\tilde #1}}
\newcommand{\bq}{\begin{eqnarray}}
\newcommand{\eq}{\end{eqnarray}}
\newcommand{\bm}[1] {\mbox{\boldmath{$#1$}}}
\newcommand{\bi}{\begin{itemize}}
\newcommand{\ei}{\end{itemize}}

\begin{document}
\title{ Quark Transverse Momentum
          Distributions
inside a nucleon : a Light-Front
  Hamiltonian Dynamics study
\thanks{Presented at Light Cone 2012}%
}
\author{ Emanuele Pace 
\vspace{-1mm}
\address{Universit\`a di Roma ``Tor Vergata'' and INFN, Roma 2, Italy }
\\
\vspace{4mm}
Giovanni Salme'
\vspace{-1mm}
\address{INFN Sezione di Roma, Italy}
\\
\vspace{4mm}
Sergio Scopetta
\vspace{-1mm}
\address{Universit\`a di Perugia and
 INFN, Sezione di Perugia, Italy}
\\
\vspace{4mm}
Alessio Del Dotto
\vspace{-1mm}
\address{Universit\`a di Roma Tre and INFN, Roma 3, Italy}
}
\maketitle
\begin{abstract}
Through an impulse
approximation analysis of single spin Sivers and Collins asymmetries
in the Bjorken limit, the possibility to extract the quark transverse-momentum distributions in 
the neutron from
semi-inclusive deep inelastic electron scattering off polarized $^3$He is illustrated.
 The analysis is generalized to 
  finite  momentum transfers in a light-front Poincar\'e
covariant framework, defining the light-front spin-dependent spectral function of a J=1/2
system.
The definition of the light-front spin-dependent spectral function for constituent quarks
in the nucleon allows us to show that, within the
light-front dynamics, only three of the six leading twist T-even 
transverse-momentum distributions 
are independent.
\end{abstract}
  
\section{Introduction}
The interest in the  transverse momentum-dependent parton
distributions (TMDs) inside a nucleon is a consequence of the
"spin crisis": most of the proton spin is carried
 by the quark orbital angular momentum, $L_q$, and by the gluons.
Information on the TMDs \cite{Barone} and then on $L_q$
can be  accessed through non forward processes, as 
semi-inclusive deep inelastic electron scattering (SIDIS). 
In  particular single spin asymmetries {{SSAs}} in the scattering of an unpolarized 
electron beam
on a transversely polarized target allow one to 
distinguish the Sivers and the 
Collins asymmetries, which can be
 expressed in terms of different 
TMDs
and {{fragmentation functions}} (ff) \cite{Barone,Sco}.
Actually a large Sivers asymmetry was measured in the process ${\vec p}(e,e'\pi)x$ \cite{Hermes}, 
while only a
small Sivers asymmetry was measured in ${\vec D}(e,e'\pi)x$ \cite{COMPASS}. 
Then one can infer a strong
flavour dependence of the asymmetries,
 confirmed by recent data \cite{Lussino}.
This puzzle has attracted a great interest in obtaining new information on the neutron TMDs.

In Ref. \cite{Cates}  
 the possibility was proposed
to extract information on the neutron TMDs from 
experimental measurements of the single spin asymmetries  on $^3$He. In
Ref. \cite{Sco} an impulse approximation (IA) approach to SIDIS off $^3$He in the Bjorken limit was
presented and applied to demonstrate how neutron Sivers and Collins SSAs can be extracted from the
$^3$He and proton ones.


In this contribution we develop a light-front approach to SIDIS on $^3$He, already considered in
\cite{Dotto}, to take care of relativistic effects at finite values of the momentum transfer 
through a light-front spectral function for $^3$He.

Eventually, through the definition of the light-front spin-dependent spectral function 
 for a nucleon, considered as a system of three quarks, we show that  within the
light-front dynamics only three of the six leading twist time-reversal-even 
transverse-momentum distributions 
are independent.

\section{Neutron single-spin asymmetries and a polarized $^3$He target}
As is well known, a polarized $^3$He is an ideal target 
to study the {{neutron}}, since at a 90\% level a polarized $^3$He is equivalent to
a polarized neutron. In order to take care of the motion of the bound nucleons in $^3$He,
dynamical nuclear effects in inclusive deep inelastic electron scattering  processes,
 {{$^3\vec{He}( e,e')X$}}, (DIS) were evaluated with a realistic 
spin-dependent spectral function  (SDSF)
for $^3{{\vec{He}}}$
\cite{Ciofi}. It was found  
  that the formula
  \vskip -0.4cm
\begin{equation}
  {{A_n }}\simeq {1 \over 
{{p_n}} f_n} \left 
( {A^{exp}_3} - 2 
{p_p} f_p
{{A^{exp}_p}} \right )~, \quad   \nonumber \\
{ \quad \quad \quad \quad 
(f_p, f_n \quad {{dilution factors}})} 
\end{equation}
\vskip -0.1cm
can be safely adopted to extract the neutron information from $^3$He and proton data. 
Actually this formula is widely
 used by experimental collaborations to this end.
All 
the nuclear effects are hidden
in the proton and  neutron {{"effective polarizations"}} $p_p$ and $p_n$. In \cite{Sco} the values
$ {{p_p}} = -0.023  $, 
${{p_n}}= 0.878 $ were obtained for the AV18 interaction \cite{AV18} using the overlaps calculated
in \cite{Pace}. Only negligible effects can be found including three-body interactions.

To investigate if an analogous formula can be used to extract the {{SSAs}},
  in \cite{Sco}
the process {{$^3\vec{He}( e,e'\pi)X$}} 
was evaluated  in the Bjorken limit and
in impulse approximation, i.e.
 the  final state interaction (FSI) was considered only between
 the  two-nucleon system
 which recoils, while no interaction was considered between
the measured fast, {{ultrarelativistic}}  ${{\pi}}$ 
and   the remnant.
 In IA, {{SSAs}} for $^3He$
involve convolutions of SDSF for $^3He$
 with TMDs
 and fragmentation functions. 
The {{nuclear} effects} on 
 ff
are new with respect to the {DIS} case.
Ingredients of the calculations were: i)
a realistic SDSF
for {{$^3$He}} \cite{Kiev,Pace},
obtained
using the {{AV18}} interaction
 and the {wave functions} evaluated by the {Pisa} group \cite{Pisa};
ii) parametrizations of data for TMDs and 
ff, whenever available;
iii) models for the unknown TMDs and 
ff.
As shown in Fig. 1,
in the Bjorken limit the extraction procedure through the  formula 
successful in DIS  works nicely 
for the Collins SSA as well, replacing in Eq. (1) $A^{exp}_3$ with the calculated
$^3He$ asymmetry, $A^{calc}_3$, and $A^{exp}_p$ 
with the proton asymmetry corresponding to the adopted model, $A^{model}_p$. 
The same was shown to occur for the Sivers SSA \cite{Sco}.
\begin{figure}
\includegraphics[width=6.3cm]{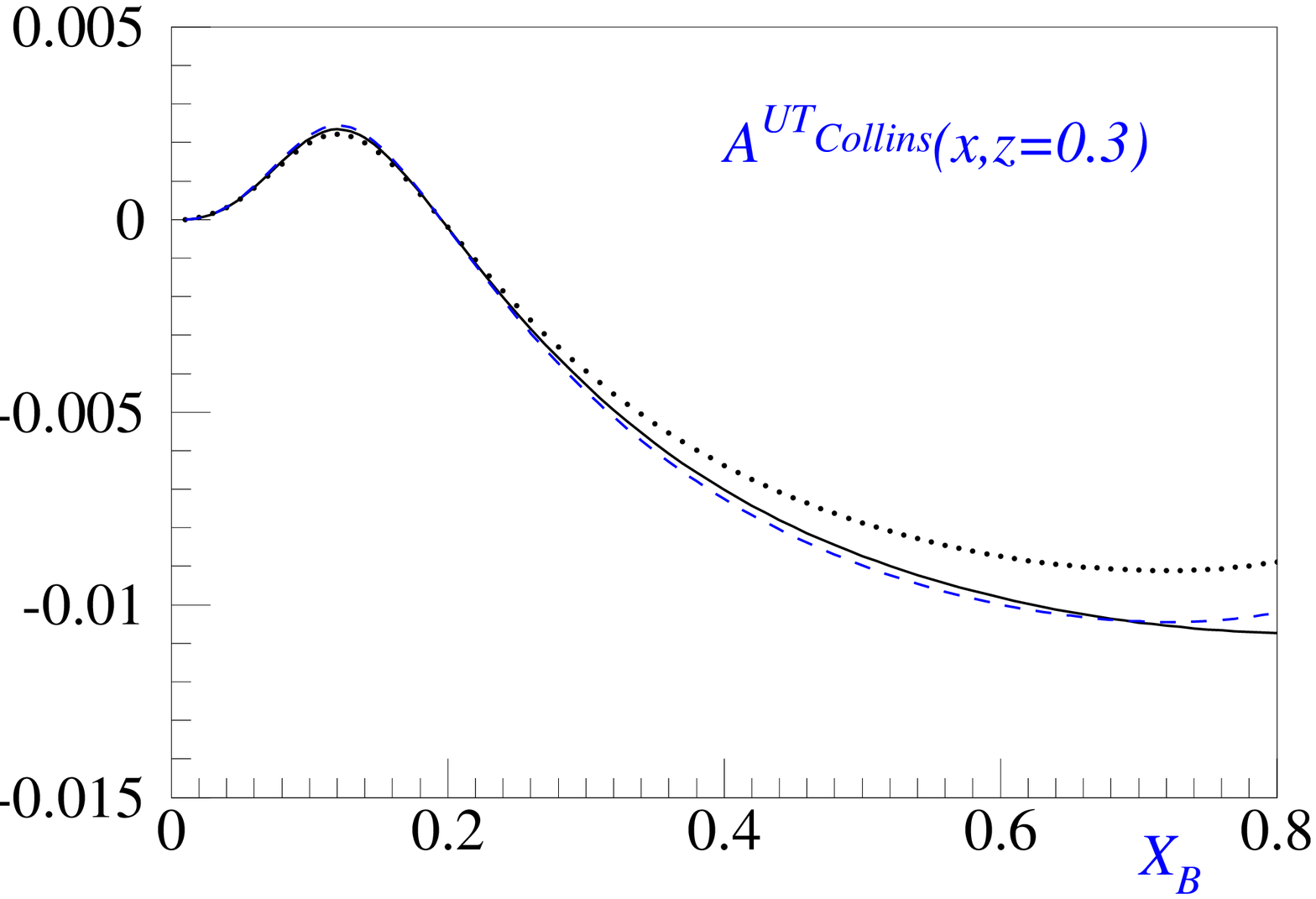}
\includegraphics[width=6.3cm]{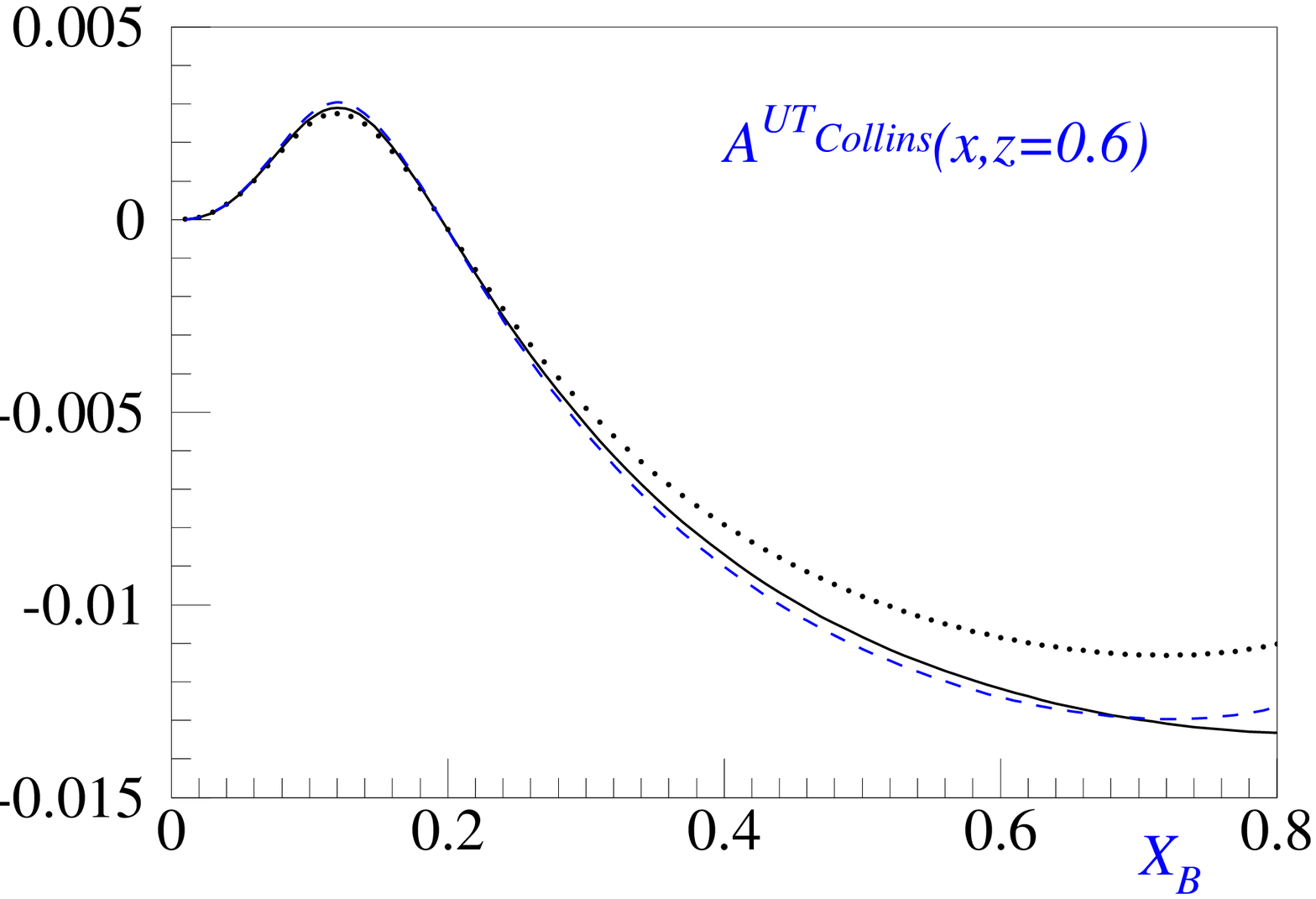}
\vspace{-4cm}
\caption{Collins asymmetry. Full line: neutron asymmetry of the adopted model, $A^{model}_n$ ;
dotted line: neutron asymmetry extracted 
from the calculated $^3He$ asymmetry neglecting the proton polarization contribution:
 $ {{\bar{A}_{n}}} \simeq {1 \over 
{f_n}} {{A^{calc}_3}}$;
dashed line:
neutron asymmetry extracted 
from the calculated $^3He$ asymmetry taking 
into account  
nuclear structure  through Eq. (1) (after \cite{Sco}).}
\label{fig:1}       
\end{figure}

In \cite{Sco} the calculation was performed in the Bjorken limit.
To study relativistic effects in the 
actual experimental kinematics, in Ref. \cite{Dotto} we adopted  the light-front (LF) form of 
Relativistic Hamiltonian Dynamics (RHD) introduced by 
Dirac.
Indeed the {{RHD} of an interacting system with a {\em{fixed number}} of on-mass-shell constituents, 
{\it plus} the Bakamijan-Thomas 
construction of the Poincar\'e generators 
allow one to generate a description of SIDIS off $^3$He
which 
is fully Poincar\'e covariant.

 
Within the LF Hamiltonian dynamics one has 
{7 kinematical generators}, 
a subgroup structure of the LF boosts, 
a separation of the {intrinsic  and the center of mass  motion} and a
{{meaningful Fock expansion}}. 

In IA the {LF hadronic tensor} for the $^3$He nucleus is:
 \vskip -0.4cm
\begin{eqnarray}
{{ 
{\cal W}^{\mu\nu}(Q^2,x_B,z,\tau, \hat{\bf h},S_{He})}} 
 \propto
 \sum_{\sigma,\sigma'}\sum_{\tau} 
 \left.\sumint \right._{\epsilon^{min}_S}^{\epsilon^{max}_S}{~d
\epsilon_{S} }
\int_{M^2_N}^{(M_X-M_S)^2} dM^2_f 
 \quad \quad
\\
 \times ~~
\int_{\xi_{low}}^{\xi_{up}} {d\xi\over \xi^2 (1-\xi)(2\pi)^3}\int_{P^{m}_\perp}^{P^{M}_\perp}{d
P_\perp\over sin\theta   }~ (P^++q^+- h^+)
~ 
{{
w^{\mu\nu}_{\sigma\sigma'}\left(\tau,{\blf q},{\blf h},{\blf P}\right)
}}\nonumber \\
\times ~{{
{\cal P}^{\tau}_{\sigma'\sigma}({\bf k},\epsilon_S,S_{He})
}}
\nonumber
\end{eqnarray}
where $ \tilde{\bf v} = \{v^+=v^0+v^3, {\bf v_{\perp} } \}$,
{$ w^{\mu\nu}_{\sigma\sigma'}
\left(\tau,{\blf q},{\blf h},{\blf P}\right)$
is the nucleon hadronic tensor} and
{{
$
{\cal P}^{\tau}_{\sigma'\sigma}({\bf k},\epsilon_S,S_{He})
$
 the {{LF}} spectral function for $^3$He
given in terms of the unitary Melosh Rotations, 
$
{{D^{{1 \over 2}} [{\cal R}_M ({\bf k})]}}$,
}}
and of the {{instant-form spectral function}}
$
{{
{\cal S}^{\tau}_{\sigma'_1\sigma_1}({\bf k},\epsilon_{S},S_{He})
}}$ \cite{Pace1}:
\bq
{ {
  {\cal P}^{\tau}_{\sigma'\sigma}({\bf k},\epsilon_{S},S_{He})
}}
\propto
\sum_{\sigma_1 \sigma'_1} 
{{D^{{1 \over 2}} [{\cal R}_M^\dagger ({\bf
k})]_{\sigma'\sigma'_1}}}~
{{
{\cal S}^{\tau}_{\sigma'_1\sigma_1}({\bf k},\epsilon_{S},S_{He})
}} ~
{{D^{{1 \over 2}} [{\cal R}_M ({\bf k})]_{\sigma_1\sigma}}}
\eq
 \vskip -0.2cm
Notice that
$
{\cal S}^{\tau}_{\sigma'_1\sigma_1}
$
is given in terms of {{three independent functions}},
${{B^{\tau}_{0,He},B^{\tau}_{1,He},B^{\tau}_{2,He}}}$ \cite{Kiev}
\bq
{{
{\cal S}^{\tau}_{\sigma'_1\sigma_1}({\blf k},\epsilon_{S},S_{He})
}}
& = & 
~\left[ 
{{
B_{0,{He}}^{\tau}(|{\bf k}|,E)
}}
~+~
{\bm \sigma} \cdot {\bf f}^{\tau}_{S_{He}}
({\bf k},E) \right]_{\sigma'_1\sigma_1}~
\eq
\vspace{-1mm}
with
$\,\,\,\,\,\,\,\,\,\,
{\bf f}^{\tau}_{S_{He}}({\bf k},E) ~=~
{\bf S}_{He}~
{{
B_{1,{He}}^{\tau}(|{\bf k}|,E)}}~+~\hat{k}~(\hat{k}
\cdot {\bf S}_A)~ 
{{
B_{2,{He}}^{\tau}(|{\bf k}|,E)
}}
$ and ${\bf S}_{He}$ the $^3$He polarization vector .

We are now  evaluating the SSAs using the {{LF hadronic tensor}}
at finite values of $Q^2$ (Eq. (2)). 
%
The preliminary results are quite encouraging. Indeed, as shown in the Table,
{{LF}} longitudinal and transverse polarizations only weakly differ and the differences 
with respect
to the non-relativistic results are small. Furthermore,  
we find that
in the Bjorken limit the extraction procedure works well within 
{{the LF approach}} as it does in the non-relativistic case.
The effect of the finite {{integration limits}}
in the actual JLAB kinematics \cite{Qian}, instead of the ones in the Bjorken limit,
is small and
 will be even smaller
in the JLAB planned experiments at 12 GeV \cite{Cates}.

Concerning the FSI, we plan to include the FSI
between the jet produced from the hadronizing quark
 and the two-nucleon recoiling system through a Glauber approach \cite{Kaptari}.
\vskip 0.2cm
\begin{center}
\hspace{-.2cm} 
\begin{tabular}{|c|c|c|c|c|c|}
\hline  &\hspace{-1mm}$proton \, {NR}$\hspace{-2mm}&\hspace{-1mm}$proton \,
{LF}$\hspace{-3mm}&\hspace{-1mm}$neutron \, {NR}$\hspace{-3mm}&\hspace{-1mm}$neutron \, {LF}$\hspace{-1mm} \\ 
\hline 
\hspace{-2mm} $\int dE d\vec{p}\,\frac{1}{2}Tr( {\cal{P}} \sigma_{z})_{\vec{S}_A=
\widehat{z}}$ \hspace{-2mm} & -0.02263 & {{-0.02231}} & 0.87805 & 
{{0.87248}} \\ 
\hline 
\hspace{-2mm} $\int dE d\vec{p}\,\frac{1}{2}Tr( {\cal{P}} \sigma_{y})_{\vec{S}_A=
\widehat{y}}$ \hspace{-2mm}
& -0.02263 & {{-0.02268}} & 0.87805 & {{0.87494}} \\ 
\hline 
\end{tabular} 
\end{center}
\section{The $J =1/2$ {{LF}} spectral function and the nucleon {{LF}} TMDs}
The TMDs for a $J =1/2$ system are introduced
through the {{q-q correlator}} (with $\psi_{\alpha}$ the $\alpha$ Dirac component 
of the quark field) \cite{Barone}
\begin{eqnarray}
  {{\Phi_{\alpha\beta}(k, P, S)}} =
   \int {d^4z} ~ e^{i k{\cdot}z}
 \langle P S | ~\bar \psi_{\beta}
(0) ~  \psi_{\alpha} (z) | P S
\rangle 
\end{eqnarray}
\begin{eqnarray} 
 \Phi(k, P, S)  = \frac12
  \left\{ \phantom{\frac1M} \hspace{-5mm}
    {{A_1}} 
\, {P}\hspace{-2mm} / \hspace{1mm} +
    { {A_{2}}} 
\, S_L \, \gamma_5 \, {P}\hspace{-2mm} / \hspace{1mm} +
  { {A_3}} 
\, {P}\hspace{-2mm} / \, \gamma_5 \, {S}_\perp\hspace{-4mm} / ~~+   \frac1{M} \, 
{{\widetilde{A}_1}} 
\, \vec{k}_\perp{\cdot}\vec{S}_\perp \,
    \gamma_5 {P}\hspace{-2mm} /
  \hspace{1mm}  + \,
  \right.
  \nonumber
\\
   \null +
  \left.
{{\widetilde{A}_2}} 
\, \frac{S_L}{M} \,
    {P}\hspace{-2mm} / \, \gamma_5 \, {k}_\perp\hspace{-4mm} /
\hspace{2mm} +  \, \frac1{M^2} \, 
{{\widetilde{A}_3}} 
\, \vec{k}_\perp{\cdot}\vec{S}_\perp \,
    {P}\hspace{-2mm} / \, \gamma_5 \, {k}_\perp\hspace{-4mm} / \hspace{2mm}
  \right\}~,~
\end{eqnarray}
so that the {{six twist-2 T-even TMDs}}, {{$A_i,
\widetilde{A}_i ~ (i=1,3)$ }}, can be obtained by proper traces of 
{{$\Phi(k, P, S)$}}.
 Let us consider the  contribution to the {{correlation function}}
from  on-mass-shell fermions
\vskip -0.6cm
\bq
{ {~\Phi_p(k,P,S)=~{(~{  k \hspace{-2mm} /}_{on}~ + ~m )\over 2 m}~
{{\Phi(k,P,S)}}~{(~{  k\hspace{-2mm} /}_{on}~ + ~m )\over 2 m} }}=
\\   = 
\sum_\sigma \sum_{\sigma'}~u_{LF}({\tilde k},\sigma')~\bar{u}_{LF}( {\tilde k},\sigma')
~{{\Phi(k,P,S)}}
~u_{LF}({\tilde k},\sigma)\bar{u}_{LF}( {\tilde k},\sigma) \nonumber
\eq 
\vskip -0.3cm
and let us identify $\bar{u}_{LF}( {\tilde k},\sigma')
~\Phi(k,P,S)
~u_{LF}({\tilde k},\sigma)$ 
with 
the {{LF nucleon spectral function}},
${\cal P}^{}_{\sigma'\sigma}(\tilde{\bf k},\epsilon_S,S)$.
The off-mass-shell minus component {$k^-$} of the struck quark is related to 
the spectator diquark energy {{$\epsilon_S $}} \cite{Pace1}.
The  
 traces of $~[\gamma^+ ~\Phi_p(k,P,S)]$,~ $[\gamma^+ ~\gamma_5~~\Phi_p(k,P,S)]$, and
 $[{k}\hspace{-2mm} / _\perp\gamma^+
\gamma_5~\Phi_p(k,P,S)]$ can be obtained in terms of the TMD's, {{$A_i,~
\widetilde{A}_i ~ (i=1,3)$}}, through Eq. (5). 
However these same traces can be also expressed through the {{LF nucleon spectral function}},
since
\bq
{ {{1 \over 2 P^+}~ Tr\left[ \gamma^+ ~\Phi_p(k,P,S) \right]}} ~ 
=
~{k^+\over 2m P^+} ~  { {Tr\left[ {\cal P}^{}_{}(\tilde{\bf k},\epsilon_S,S)
\right]}}
\eq
\vspace{-6mm}
\bq
{ {{1 \over 2 P^+}~Tr\left[ \gamma^+ ~\gamma_5~~\Phi_p(k,P,S)\right]}}=
 ~{k^+\over {2m P^+}} ~  { {
Tr\left[ \sigma_z ~ {\cal P}^{}(\tilde{\bf k},\epsilon_S,S) \right]}}
\eq
\vspace{-6mm}
\bq
{ {{1 \over 2 P^+}
Tr\left[ {k}\hspace{-2mm} / _\perp\gamma^+
\gamma_5~\Phi_p(k,P,S)\right]}} 
={k^+\over {2m P^+}} ~  {{
Tr\left[ {\bm k}_\perp \cdot {\bm \sigma} ~ 
{\cal P}^{}(\tilde{\bf k},\epsilon_S,S) \right]}} \quad \quad
\eq
In turn the traces 
${1 \over 2} {\Tr}( {\cal P} I)$, 
${1 \over 2} {\Tr}( {\cal P} \sigma_z )$, 
${1 \over 2}
{\Tr}(  {\cal P} \sigma_i)$ ($i=x,y$) can be expressed in terms of known
kinematical factors and three scalar functions,
${{B_{0,N},B_{1,N},B_{2,N}}}$, 
since the nucleon spectral function is given in terms of these
functions, in analogy withe the
$^3$He case.

Then in the {{LF approach}} with a {\em{fixed number}} of particles the six 
leading twist TMDs, $A_i,
\widetilde{A}_i ~ (i=1,3)$, 
can be expressed in terms of the previous three independent scalar functions. 
\vspace{-2mm}

\section{Conclusion}
An  realistic study in impulse approximation of the process {{ $^3\vec{He}(e,e'\pi)X$ }}
in the Bjorken limit has been described. 
The effect of nuclear  structure
in the extraction of the {{neutron}} asymmetries was 
found to be {{under control}}.
The extraction procedure of the neutron 
information from $^3\vec{He}(e,e'\pi)X$ experiments at finite $Q^2$  is now being studied
with a {LF} spectral function.

From general properties 
{{within LF dynamics with a {\em{fixed number}} of degrees of freedom}},
a few relations are obtained
among the six leading twist T-even TMDs, 
 so that {{only three}}
  of the six {{T-even TMDs are independent}}.}
 These novel relations are precisely predicted 
within {{LF}} dynamics, 
and could be experimentally
checked to test the {{LF}} description of SIDIS, at least in the valence region.



\end{document}